***Electron-phonon interaction and surface effects on the optoelectronic properties of diamondoids: A comparative study***

***F. Marsusi****

*Department of Physics, Amirkabir University of Technology, P.O. Box 15875-4413, Tehran, Iran*

***M. H. Khodabandeh***

*Iranian National Center for Laser Science and Technology, Tehran, Iran*

Unusual optoelectronic properties of diamondoids produce some discrepancies between experiments and the outstanding many-body calculation outputs. Therefore, many theoretical efforts are attracted to resolve these inconsistencies. Here first, by combining time-dependent density functional theory (TD-DFT) and Franck-Condon (FC) approximation, the effect of electron-phonon (e-ph) interaction on the optical gap (OG) of the smallest diamondoids and one of its derivatives is studied. Then, the surface effects on the e-ph coupling and the optical properties of these structures in a comparative manner are considered. We show that the collective motion of carbons modifies the previous OG of adamantane predicted by TD-DFT technique. The introduction of this effect can also fully explain the overestimated gap predicted by the diffusion quantum Monte-Carlo (DMC) method. In addition, we show that the chemistry of the surface is another noticeable effect that can influence the OG renormalization and the spectral lineshape of the system.

## I. INTRODUCTION

Nuclear motion happens on a slow time scale, compared to the electronic response. Accordingly, the electron-phonon (e-ph) interaction often has only a negligible contribution to the electronic excitation energy of the most materials. Therefore, computing static vertical excitation (see Fig. 1a) at the optimized geometry of the ground state is a good estimation of first optical transition. However, recent theoretical and experimental studies have reported significant redshift in the optical gaps (OGs) of some carbon allotropes, like as diamond and diamondoids due to e-ph interaction [1-3]. The strong e-ph coupling in these materials is explained by the specific properties of the carbon atom: the valence 2p orbitals are very close to the nucleus, and the carbon nucleus has a light mass [1,4].

Like other carbon allotropes, diamondoids illustrate remarkable chemical and physical properties and therefore have received a great attention in recent years from both theoretical and experimental points of view [5-10]. They are described by the chemical formula of $C_nH_{n+6}$ and hydrogen-terminated surface, while carbon atoms are $sp^3$ bonded to form each cage [11-13]. The smallest member of the diamondoid family is adamantane ($C_{10}H_{16}$, AD) shown Fig. 2. AD is an important molecule with many derivatives and interesting applications in multiple fields, such as industry and medicine [14]. One of AD derivatives is the 1,3,5-adamantantriol ($C_{10}H_{16}O_3$, $AD(OH)_3$), see Fig. 3. $AD(OH)_3$ is synthesized by replacing three hydrogen atoms of AD with three OH and belongs to the $C_{3v}$ point group.

A great deal of theoretical efforts including many-body techniques has been devoted to computing the OG of diamondoids [5,15]. Previous static diffusion quantum Monte Carlo (DMC) calculations [15] have overestimated the experimental AD OG [16] by approximately 0.8(1) eV, and it has been proposed to evaluate the effect of the lattice dynamics to explain this inconsistency (see Fig. 1b). On the contrary way, static calculations within the local-density approximation (LDA) in the density-functional theory (DFT) framework underestimate the experiments and static DMC by about 0.7 eV and 1.6(1) eV, respectively [15]. Static DFT and time-dependent DFT (TD-DFT) calculations using B3LYP hybrid functional [17, 18] reproduce well the experimental OGs of diamondoids [15]. B3LYP functional includes both DFT and Hartree-Fock (HF) components and therefore benefits from a cancellation of localization and delocalization errors [19]. However, from the discrepancies between this result and the static DMC outcomes, one can expect an underestimated OG of about one eV from the dynamic B3LYP and TD-B3LYP calculations.

To resolve the discrepancies between DMC outputs and experiments, the effect of the atomic motion in the photophysics of small diamondoids are studied in Ref. [1]. According to the results obtained in Ref. [1] nuclear dynamics have a significant role in optoelectronic properties of diamondoids. As such, by applying the calculated red shift due to e-ph interaction to DMC OGs the quantitative agreements with experiments is achieved.

An important subject, which has not yet been considered, is the potential role of surface phonons and the chemistry of the surface on the optoelectronic properties of carbon allotropes such as diamondoids. Nevertheless, it is not yet clear as to how the dynamical behavior adjust itself, when the surface phonons become the dominant phenomenon, and how the OG and absorption lineshape may be influenced by the chemistry of the surface?

To answer the above questions, we inspect the potentially significant role that the surface phonons may play in the absorption edge and lineshape of $AD(OH)_3$. The highest occupied molecular orbital (HOMO) of the AD is mostly localized around C-C bonds. On the other hand, the lowest unoccupied molecular orbital (LUMO) of AD wraps around the surface [7,9]. Therefore, the energy of these states are very sensitive to any change in the C-C and C-H interatomic bond lengths. As the electronic properties are influenced significantly by the energy of the HOMO and LUMU states, C-C and C-H vibronic-stretching modes of AD can affect its electronic behaviors. In addition, the energy of the HOMO and LUMO states are influenced by the corresponding frontier orbital spatial distributions. So, it would be extremely interesting to know how a modification in the spatial distribution of these orbitals will adjust dynamical behaviors of this molecule.

A modification in orbital distribution may happen with a change in the chemistry of the surface. For this purpose, here the nuclear-dynamic effects in AD and $AD(OH)_3$ are investigated. By comparing the OGs and absorption lineshapes of two structures, the effects of the surface phonons are interpreted. Our results show that the surface chemistry has strong impact on both OG and absorption lineshape. In spite of a similar backbone, carbons arrangements and orbital hybridizations, zero-point energy (ZPE) renormalizations are considerably different in the two structures.

The rest of this paper is organized as follows: In Sec. II we explain in detail our computational method within Franck-Condon (FC) approximation. In Sec. III we present and interpret our results. The effect of e-ph interaction in the OGs and lineshapes of AD and $AD(OH)_3$ are presented in a comparative manner with respect to available experimental data. Finally, a brief summary and our conclusions are given in Sec. IV.

## II. COMPUTATIONAL DETAILS

Previous study showed that scissor corrected LDA OG using FC approximation [20-22] can reproduce well both the OGs and lineshapes of small diamondoids [1]. We apply the same theoretical approach as Ref. [1] within FC approximation. At T=0, for a photon with frequency ω the absorption cross section σ(ω) in the frame of Born-Oppenheimer (BO) approximation reads:

$$\sigma(\omega)_{abs} = \frac{4\,\pi^2\omega}{3c} \sum_{\nu_0} n_{\nu_0} \sum_{j,\nu_j} \; |< \psi_{0,\nu_0}(r,Q) \mid \mu(Q) \mid \psi_{j,\nu_j}(r,Q) > |^2 \, \delta\big(E_{j,\nu} - E_{0,0} - \hbar\omega\big), \quad (1)$$

where variable $\nu_j$ stands for a vibronic state corresponds to j-th electronic level. $\psi_{0,\nu_0}$ and $\psi_{j,\nu_0}$ indicate the wave functions of the ground and j-th excited states, respectively. Within BO framework, $n_{\nu_0}$ is the population of the initial vibronic state, and $\mu$ represents the transition dipole moment between the initial and final states. Q and r stand for the nuclear and electronic coordination, respectively.

An additional approximation is done by FC principle: during an electronic transition, the time scale of electronic transition is so fast compared to the nuclear motion that we can consider the nucleus positions and velocities, and therefore μ(Q) nearly unaltered [20-22]. Within BO approximation, the molecular wave function is considered as a product of the electronic $\psi_e(Q_n)$ and nuclear $\phi_\nu$ wave functions. Electronic part of the total wave function, $\psi_e(Q_n)$, parametrically depends on the nuclear coordinates. Here $Q_n$ indicates one of 3N-6 normal coordinates describing the n-th vibrational state. By applying FC principle, the transition dipole moment term in the Eq. (1) within the BO approximation can be simplified as:

$$< \psi_{0,\nu_0}(r,Q)|\, \mu(Q)|\psi_{j,\nu_j}(r,Q) > \approx < \psi_0\, \phi_{\nu_0}|\, \mu|\psi_j\, \phi_{\nu_j} > = \mu_{0j} < \phi_{\nu_0}|\phi_{\nu_j} > . \quad (2)$$

$< \phi_{\nu_0}|\phi_{\nu_j} >$ is called overlap integral, and a transition from the ground state to an excited state is more likely to happen if the corresponding nuclear states have a large overlap. According to the FC approximation, $\mu_{0j} =< \psi_0\, |\mu|\psi_j\, >$ is supposed to be constant during the electronic transition. At T=0 the only excitation from zero vibrational-ground states is considered. For T>0, Eq. (1) must be summed over all possible initial populated states. The transition described by Eq. (1) is called an adiabatic procedure, since it also includes the structural relaxation and zero-point effects of the excited states. Due to the computational cost, FC calculations are done only for the first optical absorption path with non-zero oscillator strength. This path produces the main features of the near edge shoulder of the absorption line in which we are interested for. Ground state calculations of AD have done by B3LYP functional. In performing FC calculations, the excited-state energies and the corresponding vibrational frequencies are obtained through TD-B3LYP. Therefore, here we would be also able to obtain the overestimation of OG predicted by the static B3LYP and TD-B3LYP calculations. In the case of AD(OH)$_3$, The first dipole moment transition from HOMO (A1 symmetry) to the LUMO (A2) has a zero oscillator strength according to the parity selection rule. So, the first allowed transition couples orbitals with A1 and E symmetries, or from HOMO to LUMO+1 (E). LDA and B3LYP predict a small overlap integral between the ground- and excited-vibronic states of AD(OH)$_3$. This can stem from the large displacements predicted by these functionals between the initial and the final states. This result may be a consequence of almost non-local feature of LUMO and LUMO+1 orbitals, and also significant orbital polarization with large spatial distribution shown in Fig. 3. In such cases, long-range corrected CAM-B3LYP should improve the description of the tails of the wave functions. [23]

Both ground and excited-state calculations are done by using GAMMES program package [24-25]. 6-31G basis sets are augmented with diffuse functions in order to increase the accuracy within delocalized LUMO states [26]. The polarizations of the atoms are described by adding higher angular momentum basis functions (p,d) for hydrogen and carbons. OGs are calculated from the first non-zero vibronic transition between the ground and the relaxed-excited states. FC calculations are done by using FCCLASS code [27].

## III. RESULTS AND DISCUSSION

Applying nuclear dynamics into TD-DFT calculations influences both absorption edges and spectral lineshapes of AD and AD(OH)$_3$. However, in going from AD to AD(OH)$_3$, these features are substantially affected by the chemistry of the surface. Static and dynamic TD-DFT spectra of AD and AD(OH)$_3$ are compared in Figs. 4 and 6. The first interesting result is the different-OG renormalizations stems from nuclear motion in each structure. We obtained 700 meV redshift of absorption edge for AD due to nuclear motion, in which the ZPE contribution is about 282 meV. This result is close to the red shift of 650 meV and ZPE contribution of 320 meV predicted by the Ref. 1 using LDA functional. By applying 700 meV redshift, the DMC OG reduces from 7.35(8) [9] into 6.65(8) eV, which reproduced well the experimental value of 6.49 eV [18] within the DMC error bar. This result strongly approves the accuracy of the DMC method and its ability to predict optoelectronic properties of materials.

On the other hand, renormalized gap results from the nuclear motion of $AD(OH)_3$ represents the completely different feature. According to our outcomes, the absorption edge of $AD(OH)_3$ spectrum is shifted from 6.79 to 6.59 eV by 200 meV redshift with regards to the nuclear motions. In spite of the same carbon numbers in two structures, the ZPE contributes here is only 6 meV, which is much smaller than the ZPE contribution in AD. That we cannot always expect a significant ZPE effect among $sp^3$ carbons is of important consequent. This result also is an evidence of a much smaller e-ph interaction in $AD(OH)_3$, compared to AD. The reason may be explained by a comparison between orbital distributions in two structures, as depicted in Figs. 2 and 3. Oxygen atoms with large electronegativity absorb electrons so that the spatial distribution of the HOMO orbital in AD relocates itself from C-C bonds in AD towards oxygen atoms in $AD(OH)_3$. As shown in Fig. 3 in the new situation, the HOMO orbital is distributed in one side of the cage, over C-C bonds and mostly about oxygen atoms (rather than O-H bonds). The strongest vibrational mode of $AD(OH)_3$ occurs in C-O-H stretching mode, however, no charge is distributed around these bonds and instead a non-bonding property with a nodal feature over these bonds are observed. Alternatively, according to the new orbital distribution, HOMO and LUMO are affected mostly by vibrational bending modes with smaller energies. So, one can expect the weaker e-ph coupling energy in $AD(OH)_3$ and explains the smaller redshift appeared in the excitation energy of $AD(OH)_3$.

A detailed comparison between static- and dynamic- spectral lineshape reveals another important feature in terms of spectral sharp peaks results from involving phonons in the electronic excitations. Static and dynamic TD-B3LYP lineshapes of the two structures and the contribution of the individual transitions, including fundamentals and overtones or combination of two or more phonon modes are shown in Figs. 4-6. The envelope curve made by the superposition of the individual peaks forms the lineshapes of the spectra.

As shown in Figs. 4, like as LDA TD-B3LYP functional within FC approach can retrieve the main characteristic features and sharp peaks observed in the AD experimental spectrum. By applying 700 meV blueshift into the theoretical output, the 0→0 transition energy is shifted and coincide with the experiment in this figure. Consistent with Ref. 1, vibrational transitions involved in the first transition path between the ground and excited states broaden the AD spectral line in this region up to one eV. The adiabatic transitions between the two zero-point vibrational states allocate the absorption threshold. The C-C-C and C-C-H stretching modes, labeled with numbers 39 and 57 (from 3N-6=72 modes) assign for the sharp peaks with higher strength and produce the dominant shape of the spectral line. This series of peaks are at the distance of about 210 meV and indicated by the dashed-colored lines in Figs. 4. The individual contributions of the overtone or combination modes are shown in Fig. 4c. Transitions, including overtones: bending modes with lower energy of 47 and 109 meV; C-C and C-H stretching modes with the energy of 163 and 373 meV create certain peaks within a range of about 400 meV from the absorption edge. The combination of more phonon modes extends the spectrum width of the first excitation path into about one eV, as illustrated in this figure. Predicted TD-B3LYP spectrum within FC approximation, as well as the calculated spectrum in Ref. [1] are illustrated in Fig. 5. From this figure, the main features are observed in the AD absorption line is reproduced well by the theory is explained by the Eq. (1).

Again, in contrast to AD the dominant phenomenon in the absorption lineshape of $AD(OH)_3$ is not the C-H and C-C stretching modes. Here, the bending modes resulted from surface dynamics play an important role in the lower frequencies of the spectral line. The static transition line centered at 6.79 eV, depicted in solid black line in panels b and c of Fig. 6, is split into three new regions by considering nuclear dynamics. Each region has about 300-400 meV width (total width is about one eV), and includes several peaks. Single overtone and combination of two bending modes are responsible for the first sharp and high-intensity peaks at a distance of about 34 meV, shown in dashed lines in Fig. 6b. Also, C-C overtone stretching modes and their combination with bending modes create nearly low-intensity peaks at the end of this region. Single overtone C-H stretching modes along with the combination of two normal modes generate the second region with lower intensity. The third region with medium intensity is produced from C-O-H stretching modes and the combination of two or more other normal modes with higher energies. As a result, the broadening dynamics in $AD(OH)_3$ is totally different from AD, according to different electron distributions and chemistry of the surface.

## IV. CONCLUSIONS

In conclusion, e-ph interaction plays an important role in many carbon allotropes, such as diamondoids. The discrepancy in the OG value of AD obtained via the DMC and the experiment can be explained well by the nuclear-dynamic effects. In addition, the AD broaden-optical spectrum observed in the experiment is accurately simulated by TD-B3LYP approach within FC approximation. Surface effect on the el-ph interaction and optical properties are considered in a comparative study. According to our outcomes, OG, spectral lineshapes and the nature of the e-ph coupling may be affected significantly by the surface phonons and the chemistry of the surface. This explains why the spatial distribution of the HOMO and LUMO or even higher orbitals like as LUMO+1 are important factor in predicting the e-ph coupling behaviors. Since HOMO to LUMO+1 is the first optically allowed transition and this final state is concentrated on one side of the surface in $AD(OH)_3$, surface phonos of vibrational bending modes mostly control the type and intensity of the e-ph interaction in this structure. When the electron distribution being located over the atomic bonds, as in AD, stretching modes mostly govern the e-ph interaction.

## ACKNOWLEDGMENTS

The authors thankfully acknowledge the computational support provided by Dr. H. Afaride and Department of Energy Engineering and Physics at the Amirkabir University of Technology.

**References**

* marsusi@aut.ac.ir

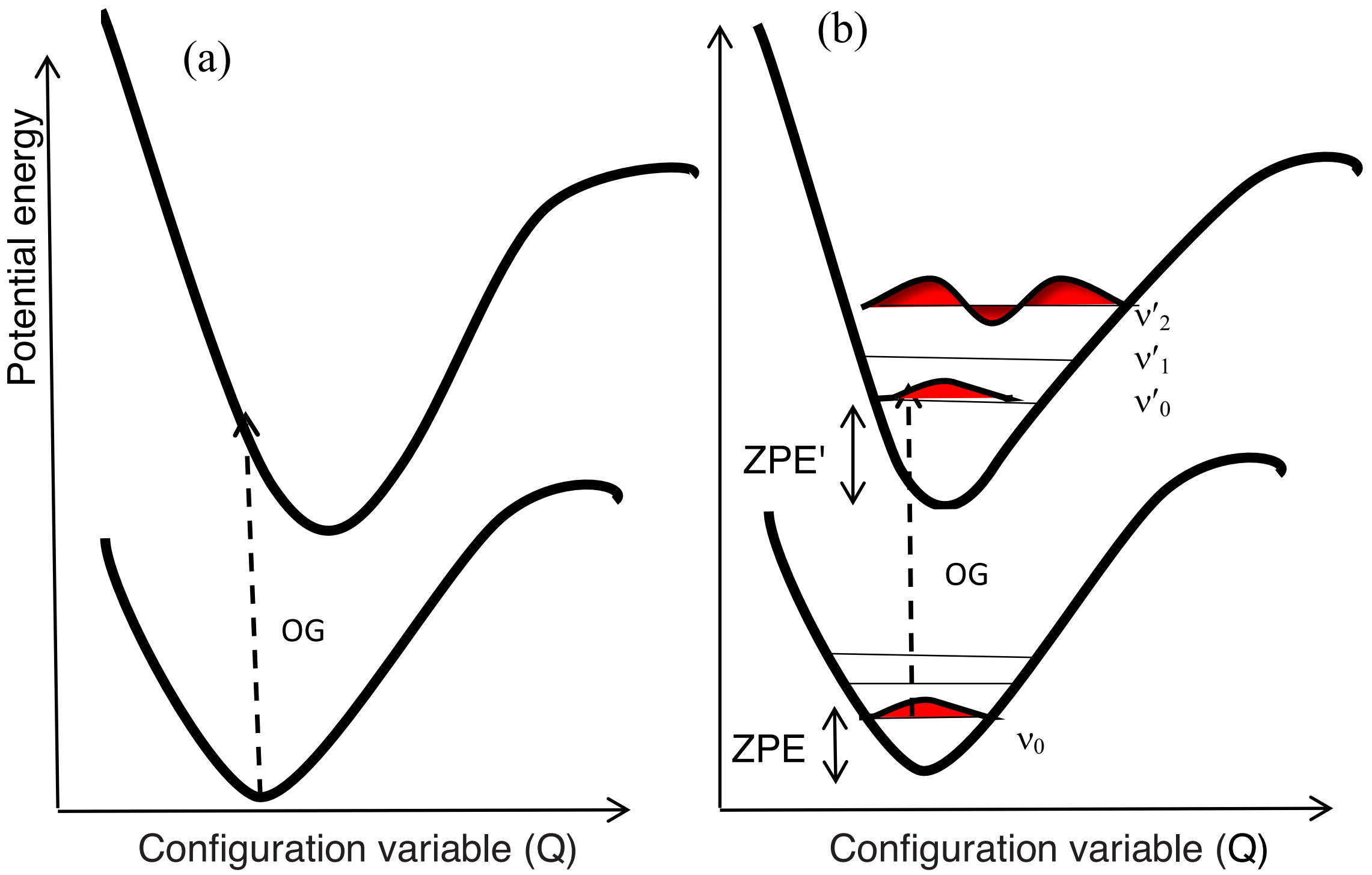

FIG. 1. (Color online) Schematic illustration of (a) static- and (b) dynamic-vertical OGs and first excitations. Excitation energy in a vertical transition is described as $E_{ex}(Q)$-$E_0(Q)$, where $E_0(Q)$ and $E_{ex}(Q)$ are the total energies of the ground and excited states at configuration geometry Q corresponding to the relaxed-ground state. Excitation energy in a dynamic transition corresponds to a transition between zero point ground state and the ν-th vibronic-excited state. The ZPE contribution is calculated by $\Delta E(ZPE) = \frac{1}{2}\sum_i h(\nu_i' - \nu_{i0})$.

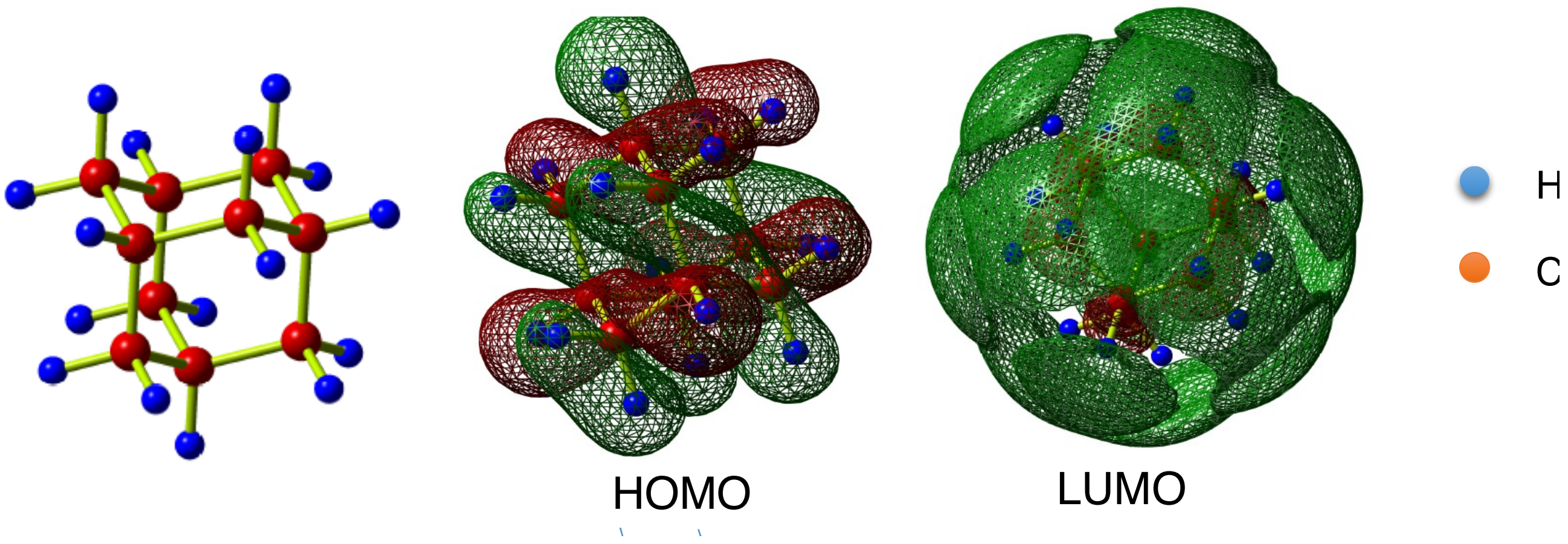

FIG.2 Model of AD structure and the corresponding isosurface of the HOMO and LUMO orbitals.

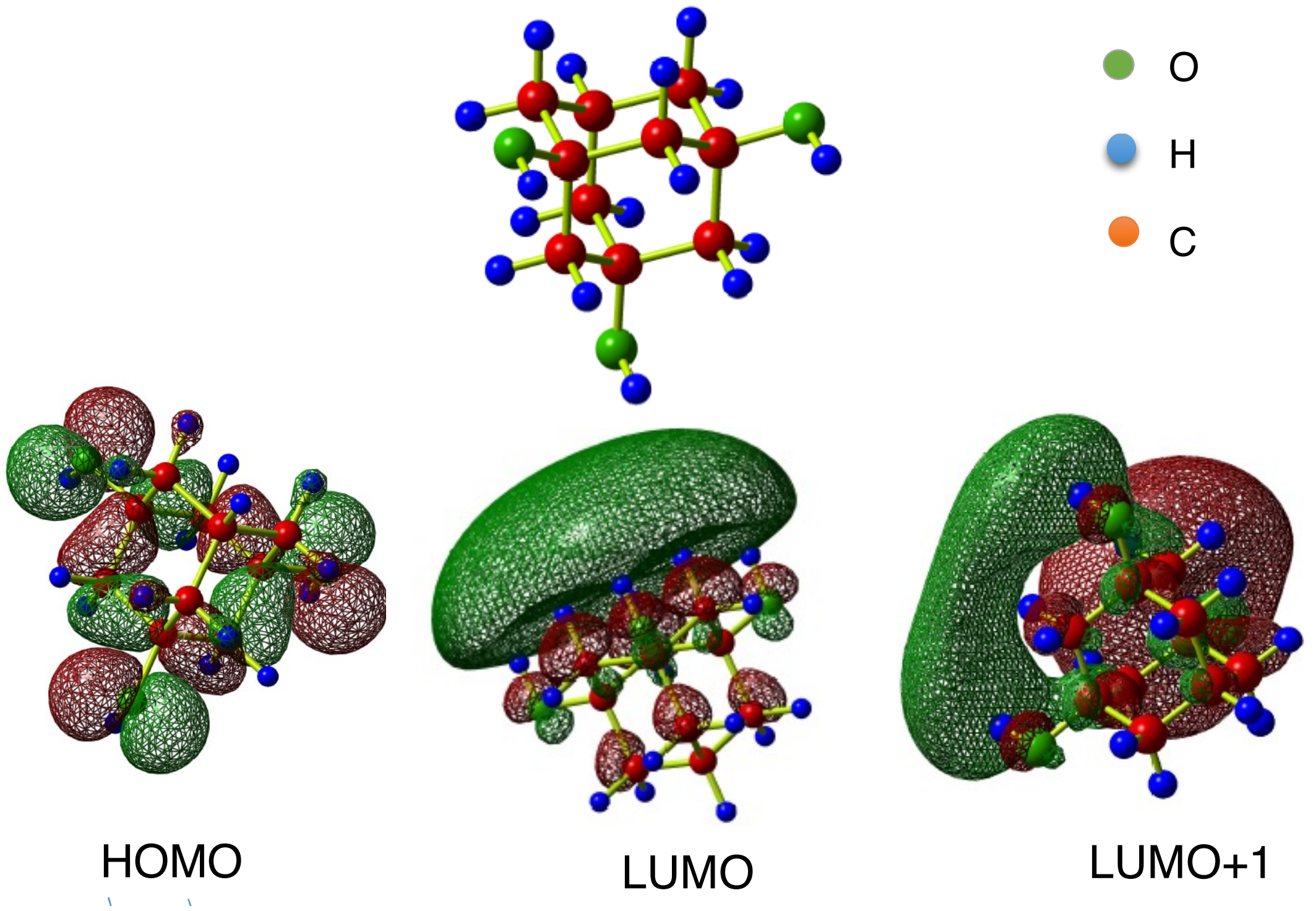

FIG. 3. (Color online) Model of $AD(OH)_3$ structure and the corresponding isosurface of the HOMO and LUMO orbitals.

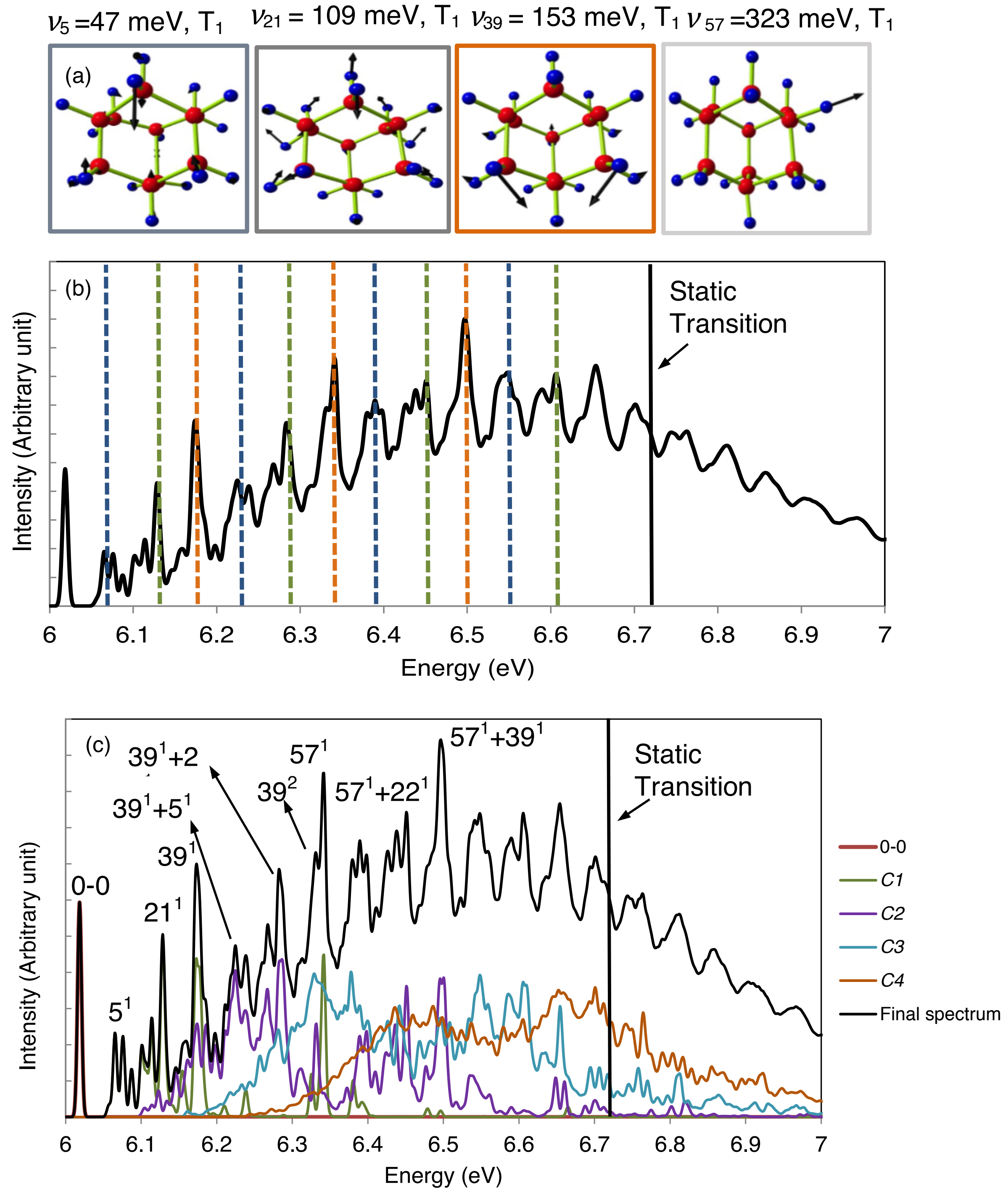

Fig. 4 (a) (Color online) Illustration of vibronic (bending, C-H and C-C stretching) which modes which significantly influence absorption lineshape of AD. (b) The static excitation energy is illustrated in solid-black line at 6.72 eV. The dynamic-excitation energy is shifted to 6.02 eV. The electronic lineshape broaden and includes sharp peaks due to nuclear-vibration effects. The sharp peaks are repeated at almost equal distances and stemmed from bending, C-H and C-C stretching modes or their combination. Each type of combination is shown in individual dashed-colored line. The smaller peaks are almost resulted from combination of stretching and bending modes. (c) Individual contributions of single-mode overtones and combinations of modes in final spectrum. $C_n$ (n=1, 2, 3, 4) represents the excitation process in which n phonon (from 3N-6 modes ) with non-zero quantum number are contributed.

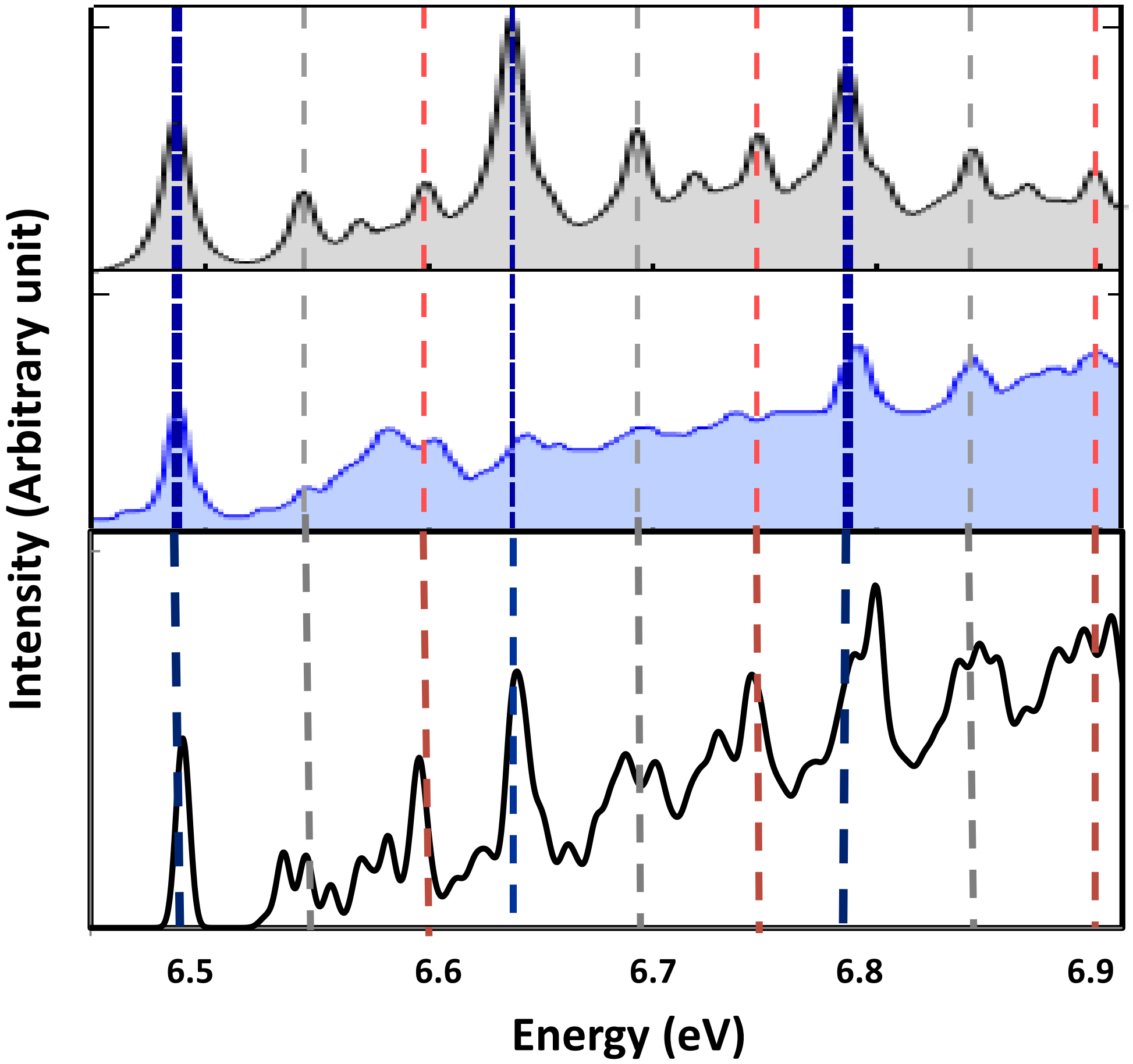

FIG. 5. (Color online) Absorption spectrum of adamantane (AD) near the absorption onset. (a) The spectrum calculated within LDA-FC framework is taken from Ref. [1]. (b) The experimental spectrum reported in Ref. [16]. (c) The spectrum calculated within FC approximation and TD-B3LYP technique using Eq. (1), while the calculated 0-0 transition energy is shifted to its experimental value.

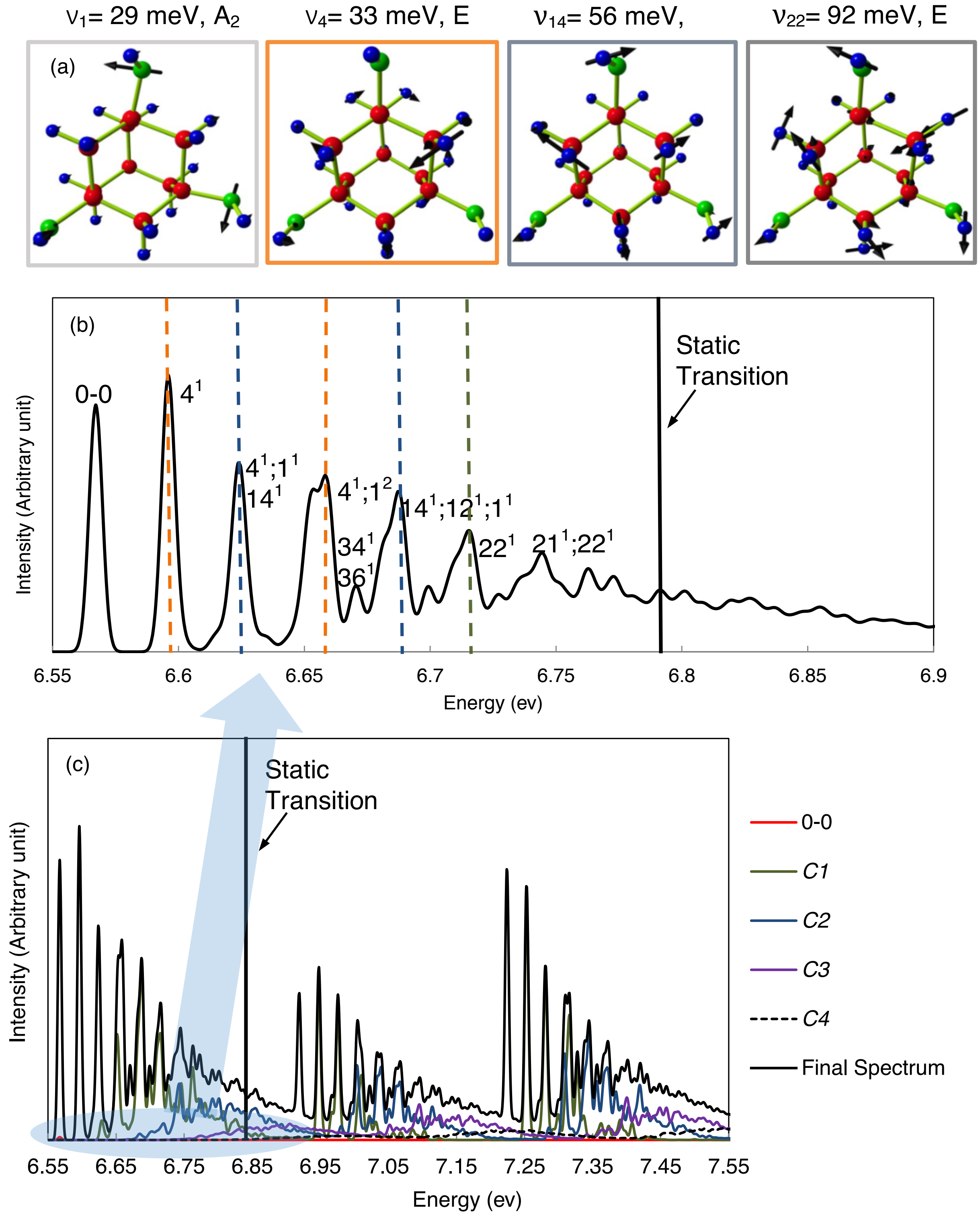

Fig 6. (a) Illustration of vibronic bending modes which significantly influence absorption lineshape of $AD(OH)_3$. (b) The static excitation energy is illustrated in black-solid line at 6.79 eV, while dynamic-excitation energy is shifted to 6.51 eV. The repeated peaks are resulted from vibronic-

bending mode at about 34 meV. The expanded lineshape includes three regions. (c) Individual contributions of single overtones and combinations of modes in final spectrum. $C_n$ (n=1, 2, 3, 4) represents the excitation process in which n phonon (from 3N-6 modes) with non-zero quantum number are contributed.